\documentstyle[preprint,revtex]{aps}
\begin{document}
\thispagestyle{empty}
\hfill\parbox{1.25in}{{\bf MAD/PH/696}\\June 1992}\par
\vspace{.5in}
\begin{title}
\bf Collider discovery limits for supersymmetric Higgs bosons
\end{title}
\author{\mbox{V.~Barger\rlap,$^*$  Kingman Cheung\rlap,$^*$
 R.~J.~N.~Phillips$^\dagger$ and A.~L.~Stange$^*$}}
\begin{instit}
$^*$Physics Department, University of Wisconsin, Madison, WI 53706, USA\\
$^\dagger$Rutherford Appleton Laboratory, Chilton, Didcot, Oxon, OX11 0QX
England
\end{instit}

\begin{abstract} \nonum\section{abstract}
The prospects for discovery
of the five Higgs bosons of the minimal supersymmetric standard model
are assessed for existing and planned future colliders, including
LEP\,I, LEP\,II, LHC and SSC. As a benchmark for comparisons, we take
a top-quark mass $m_t= 150\,$GeV and squark mass parameter $\tilde m=
1\,$TeV in evaluating one-loop radiative corrections; some results for
other $m_t$ values are also given. Searches based on the most
promising production and decay channels are taken into account.
For large regions in parameter space, detectable signals are predicted for one
or more of the Higgs bosons, but there remains a region for which no signals
would be visible at the above colliders.
\end{abstract}

\newpage
\section{INTRODUCTION}

The nature of the electroweak symmetry-breaking (EWSB) sector is the
outstanding unresolved issue of contemporary particle physics. The
Standard Model (SM) mechanism with a single scalar doublet and one
physical Higgs boson has been subject to extensive experimental and
phenomenological scrutiny~\cite{hhguide}. Experiments at LEP\,I
\cite{aleph,delphi,L3,opal}
exclude a SM Higgs boson ($H_{\rm SM}$) with mass $m(H_{\rm SM})<
57$\,GeV~\cite{davier} and future measurements at LEP\,II are expected
to explore the range $m(H_{\rm SM}) \alt 80$~GeV~\cite{saulan}. A
higher-mass SM Higgs boson will be detectable~\cite{snowmass} at the
$pp$ supercolliders LHC (15.4\,TeV) and SSC (40\,TeV). For
intermediate masses between 80~GeV and $2M_W$, searches can utilize
the $H_{\rm SM} \to \gamma\gamma$ and $H_{\rm SM} \to ZZ^* \to 4\ell$
decay channels, with high-resolution photon and lepton detection.
Above the $H \to ZZ,WW$ thresholds, the purely leptonic decays $Z \to
\ell^+\ell^-,\ \nu\bar\nu$ and $W \to \ell\nu$ give reasonable
signatures up to masses of order 0.8--1.0\,TeV, where some other EWSB
mechanism must take effect to preserve unitarity. Thus the SM Higgs
boson is detectable throughout the range of its relevance.

The minimal supersymmetric extension \cite{hhguide} of the Standard
Model (MSSM), which solves the problem of large radiative corrections
to the EWSB Higgs boson associated with the grand unified scale, has a
richer Higgs boson spectrum with five physical states $h,\ H,\ A,\
H^\pm$. Recently it has been shown that one-loop radiative corrections
can produce important shifts in both masses and couplings of these
MSSM Higgs bosons, provided that the top quark is heavy ($m_t \sim
150$\,GeV) and the squark masses are somewhat
heavier~\cite{loopcor}. Increased phenomenological attention is now
focussed on the discovery potential for these scalar states at present
and possible future
colliders.\cite{ell,bar,kal,kz,680,baer,gunion,janot,zwirner,haber}

The purpose of the present paper is to discuss systematically the
potential of different colliders to discover these MSSM Higgs bosons,
using the most promising production and decay channels, but
restricting consideration to the presently existing or planned
colliders LEP\,I, LEP\,II, LHC and SSC. As a benchmark for comparison,
we take a top-quark mass $m_t=150$\,GeV and a common supersymmetry
(SUSY) mass scale $\tilde m=1\,$TeV, following the analysis of
Ref.~\cite{680}; some results for $m_t=100$, 120 and 200\,GeV will be
compared. Following renormalization-group arguments for no-scale
models (e.g. Ref.~\cite{zwirner}) we consider the
ratio of vacuum expectation values $\tan\beta=v_2/v_1$ to be in the
range $1\alt\tan\beta\alt m_t/m_b$. In subsequent Sections, we first
describe the MSSM parameters and then consider searches at each
collider in turn. Discovery criteria are necessarily approximate,
since they depend on assumptions about future detector performance and
achievable luminosity, so the conclusions that can be reached are
tentative; in this sense our present analysis should therefore
be viewed as indicative rather than definitive.

\section{HIGGS SECTOR OF THE MSSM}

At tree level, the masses of the five Higgs bosons in the MSSM are
determined by two parameters, conveniently chosen to be $m_A$ and
$\tan\beta$, where $m_A$ is the mass of the CP-odd neutral boson and
$\tan\beta=v_2/v_1$ is the ratio of the vacuum expectation values that
give masses to up-quarks ($v_2$) and down-quarks ($v_1$). When
one-loop radiative corrections are included, a number of additional
mass parameters enter through the loops. It is still convenient to
take $m_A$ and $\tan\beta$ (evaluated at one-loop order) as
independent parameters together with the top-quark mass $m_t$, and
squark-sector mass parameters: the latter are the soft SUSY-breaking
squark masses $m_Q,\ m_U,\ m_D$, the coefficient $\mu$ of the $H_1H_2$
mixing term in the superpotential, and the coefficients $A_t$ and
$A_b$ of trilinear $\tilde t_{\rm L} \tilde t_{\rm R} H_2$ and $\tilde
b_{\rm L} \tilde b_{\rm R} H_1$ soft SUSY-breaking terms (in the notation of
Refs.~\cite{ell,680}). In our later
quantitative analysis we choose a common SUSY mass scale
\begin{mathletters} \begin{equation}
 m_Q=m_U=m_D=\tilde m=1\rm\,TeV \label{m's}
\end{equation}
and set
\begin{equation}
 A_t=A_b=2\mu=\tilde m/2 \label{A's}
\end{equation}
such that squark mixing effects are present but not overwhelming,
following Ref.~\cite{680}.  As previously indicated, we also generally
choose the top quark mass to be
\begin{equation}
m_t = 150\,\rm GeV\;.
\end{equation}
\end{mathletters}
The qualitative features of the one-loop corrections to the Higgs boson
masses can, however, be explored in terms of $m_t$ and a common squark mass
scale $\tilde m$, neglecting the smaller effects from $A_t,\ A_b,\ \mu$
and the $b$-quark mass $m_b$.  In this approximation the mass-squared
matrix for the neutral CP-even neutral Higgs bosons is
\begin{equation}
\left( \begin{array}{r@{\qquad}c}
  m_A^2\sin^2\beta+M_Z^2\cos^2\beta &
  -\left(m_A^2+M_Z^2\right)\sin\beta\cos\beta \\
  -\left(m_A^2+M_Z^2\right)\sin\beta\cos\beta &
  m_A^2\cos^2\beta+M_Z^2\sin^2\beta+\epsilon/\sin^2\beta
\end{array} \right) \;,
\end{equation}
where the modification due to radiative corrections is given by
\begin{equation}
\epsilon = {3g^2\over8\pi^2 M_W^2} \, m_t^4
\ln\left(1+{\tilde m^2\over m_t^2}\right) \;,
\end{equation}
with $g$ the SU(2) electroweak gauge coupling. Here the corrections are
controlled by a single parameter $\epsilon$; thus a change in $\tilde m$ has
the same effect as a much smaller change in $m_t$. Diagonalization leads to
the masses
\begin{eqnarray}
m_{h,H}^2 &=& {1\over2} \left[ m_A^2+M_Z^2+\epsilon/\sin^2\beta \right]
\pm {1\over2} \biggl\{ \left[ \left( m_A^2-M_Z^2 \right) \cos2\beta
+ \epsilon/\sin^2\beta \right]^2  \nonumber \\
&& \hspace{2.85in}{}+\left( m_A^2+M_Z^2 \right)^2 \sin^22\beta \biggr\}^{1/2}
\,.
\end{eqnarray}
For $\tan\beta\ge1$ these mass eigenvalues increase monotonically
with $m_A$; the upper bound on the mass of the lighter scalar boson $h$ is
\begin{equation}
\begin{array}{rcl}
m_h^2 &<&  M_Z^2\cos^22\beta+\epsilon \;, \\
      &<&  M_Z^2 + \epsilon \;.
\end{array}
\end{equation}
For $m_t=150$\,GeV and $\tilde m=1$\,TeV, this gives $m_h<115$\,GeV.
There is a corresponding lower bound on the heavier mass $m_H$,
\begin{equation}
\begin{array}{rcl}
m_H^2 &>& \displaystyle{1\over2} \left[ M_Z^2 + \epsilon/\sin^2\beta\right]
+ {1\over2} \left[ \left( M_Z^2 - \epsilon/\sin^2\beta \right)^2
+ 4M_Z^2\epsilon \,\right]^{1/2} \;, \\
      &>& M_Z^2 + \epsilon \;.
\end{array}
\end{equation}
We note that the bounds are more stringent when $\tan\beta$ is specified.

The mixing angle $\alpha$ in the CP-even sector, determined from the
diagonalization of the above mass-squared matrix, is
\begin{equation}
\tan2\alpha = { \left( m_A^2+M_Z^2 \right) \sin2\beta \over
 \left( m_A^2-M_Z^2 \right) \cos2\beta + \epsilon/\sin^2\beta } \;.
\end{equation}
The one-loop corrections to the masses and mixing angle can be large if
$\tilde m$ is well above $m_t$ and $m_t$ is well above $M_W$. The radiative
corrections to $m_h^2$ and $M_H^2$ are proportional to $m_t^4/M_W^2$, whereas
the leading corrections to the charged Higgs mass squared are proportional
to $m_t^2$ and are considerably smaller.

Figure \ref{a} shows contour plots of  $m_h,\ m_H$ and $m_{H^\pm}$
in the $(m_A,\,\tan\beta)$ plane, for $m_t=150$\,GeV or 200\,GeV and
the SUSY parameters in Eq.~(1). The exact CP-even Higgs boson
mass limits here  are $m_h<116(149)\,{\rm GeV}<m_H$ for $m_t=150(200)$\,GeV.

The couplings of the MSSM Higgs bosons depend on $\alpha$ and $\beta$
through the following factors:
\begin{eqnarray}
&& \begin{array}{l@{\quad}c@{\quad}c@{\quad}c}
 & h & H & A \\
 t\bar t & \cos\alpha/\sin\beta & \sin\alpha/\sin\beta & \gamma_5\cot\beta \\
 b\bar b, \tau\bar\tau & \llap{$-$}\sin\alpha/\cos\beta &
 \cos\alpha/\cos\beta & \gamma_5\tan\beta \\
 WW,ZZ & \sin(\beta-\alpha) & \cos(\beta-\alpha) & 0 \\
 ZA & \cos(\beta-\alpha) & \sin(\beta-\alpha) & 0
 \end{array} \label{neutral} \\
 \nonumber \\
&& \begin{array}{l@{\quad}c}
 & H^+\\
 t\bar b & m_t\cot\beta(1-\gamma_5)+m_b\tan\beta(1+\gamma_5) \\
 \nu\bar\tau & \multicolumn{1}{r}{m_\tau\tan\beta(1+\gamma_5)} \\
 ZH^+ & (1-2\sin^2\theta_W)
 \end{array} \label{charged}
\end{eqnarray}
There is no tree-level $WZH$ vertex. In the standard convention,
$0\le\beta\le\pi/2$ and $-\pi/2\le\alpha\le0$.

The factors in Eqs.~(\ref{neutral}) and (\ref{charged}) determine the
$\alpha,\beta$ dependences of the Higgs boson production and decay
vertices that enter in the various measurable subprocesses. For a given
$\tan\beta,\ \alpha\to\beta-\pi/2$ as $m_A\to\infty$; in this limit $A,\ H$
and $H^\pm$ become approximately degenerate and physically irrelevant,
while the $h$ couplings approach those of the SM Higgs boson. Figure~\ref{c}
shows the regions where the $W,\,t,\,b$ couplings of $h$ are within 10\%
of the corresponding $H_{\rm SM}$ couplings. We see that the MSSM would be
almost indistinguishable from the SM (with $h$ in the role of $H_{\rm SM}$)
for $m_A\agt200$\,GeV and  $\tan\beta>1$ unless some of the additional
particles $A,\ H$ or $H^\pm$ could be discovered.

The decays of an off-shell $Z^*$ to $Zh$ and $Ah$ are complementary in
the sense that $\sin^2(\beta-\alpha)+\cos^2(\beta-\alpha)=1$; thus any
suppression in one coupling is accompanied by an enhancement in the
complementary coupling. Similarly, the decays of $h$ (or $H$) to $WW,\ ZZ$ and
$ZA$ are complementary. Figure~\ref{c} shows a contour plot of
$\sin(\beta-\alpha)$ in the $(m_A,\,\tan\beta)$ plane; it also shows contour
plots of the
quantities $(\cos\alpha/\sin\beta)$ and $(\sin\alpha/\sin\beta)$ that
govern the strength of $h$ and $H$ couplings to $t\bar t$ and the quantities
$(-\sin\alpha/\cos\beta)$ and $(\cos\alpha/\cos\beta)$ that govern couplings to
$b\bar b$. These five factors control the $W,\ t$ and $b$ loop contributions
for $h$ and $H$ decays into two photons; we see that they vary strongly across
the parameter space.

The partial widths for neutral Higgs to $\gamma\gamma$ transitions have
the form~\cite{hhguide}
\begin{equation}
\Gamma(X\to\gamma\gamma) = \Gamma_0(X) (\alpha/\pi)^2
\bigl| I(X\to\gamma\gamma) \bigr|^2 \;,
\end{equation}
where $I(X\to\gamma\gamma)$ represents the loop integral contributions and
\begin{equation}
 \Gamma_0(X)=G_F m_X^3 \Big/ \left(128\pi\sqrt2\,\right) \;.
\end{equation}
The $b$-loop, $t$-loop and $W$-loop contributions have the following forms:
\begin{mathletters}
\begin{eqnarray}
I(h\to\gamma\gamma) &=& {4\over9}I_b{\sin\alpha\over\cos\beta} -{16\over9}I_t
{\cos\alpha\over\sin\beta} +
7I_W\sin(\beta-\alpha) \;, \label{I(h)}\\
I(H\to\gamma\gamma) &=&-{4\over9}I_b{\cos\alpha\over\cos\beta} -{16\over9}I_t
{\sin\alpha\over\sin\beta} +
7I_W\cos(\beta-\alpha) \;, \label{I(H)}\\
I(A\to\gamma\gamma) &=& {2\over3}I'_b \tan\beta +{8\over3}I'_t\cot\beta \;.
\label{I(A)}
\end{eqnarray}
\end{mathletters}
Here the integrals $I_q,\, I_W,\, I'_q$ are real and positive for
$m_X\le2m_q,\, 2M_W,\, 2m_q$ respectively. In the regime where
$m_X<M_W,\,m_q$ the integrals all approach unity. The quantities
$-{4\over9}I_b,\ -{16\over9}I_t$ and $7I_W$ are shown as functions of
$m_X$ in Fig.~\ref{d}. For an intermediate mass $h$ or $H$, the $W$-loop
contribution dominates unless the factor $\sin(\beta-\alpha)$ or
$\cos(\beta-\alpha)$ becomes very small. Although the $b$-loop contributions
are usually negligible, they can become significant at large $\tan\beta$. These
two-photon widths may be directly measurable in the future at a $\gamma\gamma$
collider~\cite{borden};
 Fig.~\ref{e} shows their dependences on the Higgs mass for $\tan\beta=2$, 5
and 30, with the parameter choices of Eq.~(1).

The partial widths for neutral Higgs to $gg$ transitions have the
form~\cite{hhguide}
\begin{equation}
\Gamma(X\to gg) = {9\over8} \Gamma_0(X) (\alpha_s/\pi)^2
\bigl| I(X\to gg) \bigr|^2 \;,
\end{equation}
where $I(X\to gg)$ represents the loop contributions. The formulas for
$I(X\to gg)$ are obtained from those for $I(X\to\gamma\gamma)$ by
simply setting $I_W=0$ in Eqs.~(\ref{I(h)})--(\ref{I(A)}).

The cross-sections for Higgs boson production via $\gamma\gamma$ fusion
or $gg$ fusion are directly proportional to the corresponding partial
widths above.

In addition to the conventional Higgs boson decays into fermions and
gauge bosons, the MSSM allows the decays
\begin{equation}
h\to AA; \ H\to hh,\,AA,\,AZ; \ A\to Zh  \label{decays a}
\end{equation}
into other Higgs bosons, when the kinematics permits;
%Other conceivable channels such as $H^\pm\to W^\pm h$ or $A\to ZH$are
%suppressed or kinematically forbidden.
these new decays modes usually dominate over decays into fermions and weak
bosons in the regions where they are allowed, unless there is an accidental
suppression of the coupling for particular $\alpha,\,\beta$ values.
Figure~\ref{f} shows the allowed regions for these decays in the
$(m_A,\,\tan\beta)$ plane. We see that one or more of these decays are
allowed over a large part of the parameter space; in particular $H\to hh$
can occur widely,
although there is a forbidden region (shaded in Fig.~\ref{f}) and also a
suppressed band in the neighborhood of a zero of the $Hhh$ coupling
\begin{equation}
f_h = \cos2\alpha\cos(\beta+\alpha) - 2\sin2\alpha\sin(\beta+\alpha) \;,
\end{equation}
shown by a dashed curve in Fig.~\ref{f}.
The new decay modes such as $H\to hh$ lead to higher-multiplicity final
states, but if the $H$ can be produced in conjunction with $Z$, through
$e^+e^-\to Z^*\to ZH$, this is not necessarily a major liability.
In fact, the possibility that one might eventually be able to measure
such decays would offer extra tests of the Higgs sector couplings.

Figure \ref{g} illustrates the total widths of the MSSM Higgs bosons
versus mass compared to the SM value.
Alternatively, Fig.~\ref{new8} shows contour plots of these MSSM total widths
in the $(m_A,\,\tan\beta)$ plane.
Figure~\ref{h} illustrates the
branching fractions for the dominant modes and the most detectable modes.

\section{LEP\,I \lowercase{and} LEP\,II \lowercase{$e^+e^-$} COLLIDERS}

At the LEP\,I $Z$ factory with $\sqrt s\simeq M_Z$, the channels
\begin{equation}
e^+e^-\to Z\to Z^*h,\,Ah
\end{equation}
are accessible if $h$ and $A$ are sufficiently light. These channels are
complementary in the sense that the $Z^*h,\,Ah$ cross sections are
proportional to $\sin^2(\beta-\alpha),\,\cos^2(\beta-\alpha)$ respectively,
and therefore cannot be simultaneously suppressed by these coupling factors.

Comprehensive searches for these Higgs boson signals have been made at
LEP\,I, covering many decay channels ({\it e.g.}\
$h,A\to \tau^+\tau^-, \, {\rm jet + jet}, \, \mu^+\mu^-, \,
\pi^+\pi^-$)~\cite{aleph,delphi,L3,opal}. The null results of the ALEPH
searches for $Z^*h$ and $Ah$ signals~\cite{aleph} exclude the regions
of the $(m_A,\,\tan\beta)$ plane shown in Fig.~\ref{i} for our SUSY
parameter choices and $m_t=100$, 120, 150 or 200~GeV (deduced from the
corresponding ALEPH bounds~\cite{aleph} on $\sin^2(\beta-\alpha)$ and
$\cos^2(\beta-\alpha)$\,). We see that the excluded regions depend
 considerably on the value of $m_t$ (other inputs being held fixed).
Accumulating higher statistics and combining the results from all four LEP
detectors~\cite{aleph,delphi,L3,opal} will tighten these parameter bounds in
the future.

With the planned upgrade to LEP\,II
at CM energy $\sqrt s\simeq200$\,GeV, the possible MSSM Higgs production
channels will be
\begin{equation}
  e^+e^-\to Z^*\to Zh,\,Ah,\,AH \; ;
\end{equation}
$ZH$ production is either kinematically forbidden or highly suppressed for
our parameter range.
After decays these channels can yield the final states
$\ell^+\ell^-jj,\, \nu\bar\nu jj,\,\tau\bar\tau jj$ and $jjjj$
(where $j$ denotes a hadronic jet).

   Simulations of SM Higgs boson signals and backgrounds in the three
channels $e^+e^-\to \nu\bar\nu jj,\, \ell^+\ell^-jj,\, jjjj$ have been
presented for LEP\,II in Ref.~\cite{saulan}; the results show that
$H_{\rm SM}$ can be detected at least up to 80\,GeV in the $\nu\bar\nu jj$
and $\ell^+\ell^-jj$ channels and at least up to 60\,GeV in the $jjjj$
channel with integrated luminosity 500\,pb$^{-1}$ per detector.
The limited sensitivity
in the $jjjj$ channel is due to combinatorial problems and large
$e^+e^-\to W^+W^-(ZZ) \to jjjj$ backgrounds that obscure any Higgs
mass peaks near $M_W$ or $M_Z$.
These SM simulations can be rescaled approximately to estimate the
detectability of the MSSM $e^+e^-\to Zh$  signals, that  differ from
the SM signals essentially by the cross-section factor $\sin^2(\beta-\alpha)$.
Figure~\ref{j} shows the limits of detectability for the
$Zh$ signals, defined by the requirement
\begin{equation}
S\Big/\sqrt B \;=\; \rm number\ of\ signal\ events \bigg/
\sqrt{\rm number\ of\ background\ events} \;\ge\; 4 \;, \label{events}
\end{equation}
for an integrated luminosity ${\cal L}=500\,\rm pb^{-1}$, obtained by
rescaling the simulations of Ref.~\cite{saulan}. This luminosity
corresponds approximately to 2~years LEP\,II running at one intersection.
We here add the significance $S \big / \sqrt B$ of the
three  channels $Zh\to\ell^+\ell^-jj,\; \nu\bar\nu jj,\; jjjj$ in quadrature,
considering only channels containing four or more events. The four
standard deviation significance required in Eq.~(\ref{events}) implies
a higher level  of confidence for big signals (where gaussian statistics apply)
than for smaller signals (with Poisson statistics); however, requiring a
 minimum signal $S>4$~events per included channel ensures that a reasonable
level of confidence is maintained.

The $e^+e^-\to Ah,\,AH$ channels lead to $\tau\tau jj$ and $jjjj$ signals;
however the $jjjj$ simulations of Ref.~\cite{saulan} cannot be used to
estimate the acceptances and backgrounds in these cases, since the cuts
imposed include an explicit fit to an $e^+e^-\to ZH_{\rm SM}$ hypothesis.
The $\tau\tau jj$ channel was not addressed in Ref.~\cite{saulan} but has
been used at LEP\,I \cite{aleph,delphi,L3,opal} and advocated for higher energy
$e^+e^-$ colliders~\cite{janot}; it has the advantage that the background from
$e^+e^-\to ZZ$ is relatively small, compared to the $W^+W^-$ and QCD
backgrounds in the $4j$ channel, and it escapes the combinatorial problem
of the $4j$ case. The missing neutrinos from the $\tau$ decays are
approximately collinear with the observed $\tau$ decay products; the
magnitude of these two missing momenta can therefore be reconstructed
from energy-momentum constraints (there is a 1C fit, allowing for
initial-state radiation of a hard photon along a beam direction). In the
distributions of the reconstructed invariant masses $m(\tau\tau)$ and $m(jj)$,
the Higgs boson signals appear as narrow peaks with widths determined by
the experimental resolution; it is advantageous to add these $m(\tau\tau)$
and $m(jj)$ signals to improve statistics in the peaks.
The most serious  $\tau\tau jj$  background is that
from  $e^+e^-\to ZZ$;  all other backgrounds can be
reduced to insignificance by suitable cuts, as demonstrated in
Ref.~\cite{janot} for a possible future
linear collider with $\sqrt s=500$\,GeV.

In our present studies of  $\tau\tau jj$  signals
at LEP\,II, we assume that the irreducible  $ZZ$  background can
be estimated directly from the work of Ref.~\cite{janot},
rescaled to take account of the higher $ZZ$ production cross
section and lower luminosity expected at LEP\,II. This background contains
smearing from experimental resolution
and the uncertainties of tau reconstruction.   We assume that
the MSSM signals have approximately 50\% detection efficiency
(as in Ref.~\cite{janot}) and have the same smearing as the
background.  For an isolated Higgs boson peak, we define the signal strength
$S$ to be the total number of Higgs-decay counts in a 10 GeV mass bin centered
at the peak.  The background strength $B$ is taken to be the total
number of $Z$-decay counts (both from $ZZ$ and from $Zh,ZH$
production with the resolution of
Ref.~\cite{janot}) falling in the same mass bin.
When two Higgs peaks approach within 10~GeV we
combine them; the signal strength $S$ is then the
number of Higgs counts expected in a 10~GeV bin centered at the weighted mean
mass.  If the signal bin center is
separated by more than 5\,GeV from $M_Z$, our discovery criteria are
$S\Big/\sqrt B>4$ with $S>4$, for integrated
luminosity 500\,pb$^{-1}$. With such a signal, we expect that a
distinct Higgs peak will be seen or that a recognizable
distortion of the $Z$ peak will be evident.
But if the separation from $M_Z$ is less than 5~GeV, we can only infer
the presence of a new signal if the height of the supposed $Z$ peak differs
substantially from the expected $ZZ$ background contribution. In this latter
case we rely entirely on normalization and therefore require a higher degree of
significance. Here the signal $S$ is defined to be the sum of all MSSM
 ($h$, $H$, $A$ and $Z$) contributions falling in a 10~GeV bin centered at
$M_Z$, and $B$ is the expected $ZZ$ background in the same bin; in this case we
define a discoverable signal to have $S\Big/\sqrt B>6$ with $S>5$ counts.

Figure~\ref{k} shows the regions in the $(m_A,\, \tan\beta)$ plane where
a $\tau\tau jj$ Higgs signal would be detectable at LEP\,II with the
criteria above and the parameters of Eq.~(1) with $m_t=120$, 150 or
200~GeV.  The inaccessible regions at lower left occur because $h \to AA$
decay dominates here (see Figure~\ref{f}) and the $\tau\tau jj$ signals
are consequently suppressed.
Through most of the discovery region just one new peak (corresponding
to $h$ alone or to overlapping $h,A$ peaks) would be discernible;  at small
$m_A$  there are regions  where distinct  $h$  and  $A$ peaks would be
predicted, but these regions are already excluded by the LEP\,I
searches~\cite{aleph} (see Fig.~\ref{i}).

The above LEP analyses do not rely on $b$-jet identification.
The inclusion of $b$-tagging with vertex detection will expand the Higgs boson
discovery limit since the $Z$-decay background has contributions from all quark
flavors while the Higgs boson contribution is mainly from $b\bar b$.
Present $b$-tagging at LEP typically has efficiency of order 0.38 for $b\bar b$
states compared to 0.11 for the average $Z\to jj$ decays. Then if we adopt
these efficiencies and consider
for example the $Z\to\ell^+\ell^-jj$, $\nu\bar\nu jj$ and $Ah\to\tau\tau jj$
signals, where the Higgs signals all have $b\bar b$ jets and
the backgrounds come mainly from $ZZ$ production, the
significance $S\big/\sqrt B$ of a tagged Higgs signal would be enhanced above
that of the corresponding untagged signal by approximately a factor
$0.34\big/\sqrt{0.11}=1.14$, a helpful but not dramatic improvement. For the
present we conservatively neglect such developments.

Searches for charged Higgs bosons at LEP\,I~\cite{chargehiggs} have excluded
various ranges of $m_{H^+}$ below 43~GeV, but such values are well below the
expected MSSM range (see Fig.~\ref{a}). Future searches at LEP\,II will be
sensitive to $H^+H^-$ production up to perhaps 70~GeV mass (the much larger
$WW$ background making higher masses hard to detect); the corresponding MSSM
parameter region is marginal in our present discussion and appears to be
excluded already by the LEP\,I $h$ and $A$ searches (see Fig.~\ref{i}). Charged
Higgs searches at LEP are therefore not likely to affect the MSSM discovery
limits.

\section{\ \lowercase{$pp$} SUPERCOLLIDERS}

Extensive analyses have concluded that a SM Higgs boson in the intermediate
or high mass ranges can be detected at the LHC ($\sqrt s=15.4\,$TeV) and
the SSC ($\sqrt s=40\,$TeV) $pp$ supercolliders~\cite{snowmass}. However,
in the MSSM the situation may be very
different~\cite{kz,680,baer,gunion};  there are many more Higgs boson
states to consider, the couplings may differ  markedly from the SM values and
new decay channels may be present.  In the following we focus attention on the
most promising signals. Since the LHC with a factor of 10 higher luminosity
can cover about the same  ground as the SSC\cite{snowmass}, except at high
Higgs boson mass values, we illustrate with results  obtainable at the SSC. The
discussion below considers the individual  production subprocesses with decays
that yield viable signals.

\subsection{$gg\to h,\,H,\,A\to\gamma\gamma$}

The largest neutral Higgs boson production cross sections are due to
gluon-gluon fusion, and for Higgs masses in the intermediate range
the $\gamma\gamma$ decay mode gives a viable signal provided that excellent
$m(\gamma\gamma)$ mass resolution can be achieved.
The proposed GEM detector~\cite{gem} at the SSC is capable of mass resolution
$\Delta m/m(\gamma\gamma)\le 1\%$, that gives sufficient background
suppression to detect the inclusive $H_{\rm SM}\to\gamma\gamma$ signal
through the mass range 80--160\,GeV. We use the GEM simulations of the
background with a liquid argon  calorimeter and their minimal 55\% signal
efficiency after cuts~\cite{gem} to determine the significance $S\Big/\sqrt{B}$
of signal versus background in $\pm 1$\%
mass intervals at the Higgs boson masses ({\it i.e.}\ a bin width of twice the
expected experimental resolution of 1\%). Figure~\ref{l} gives the raw
$pp\to gg\to h,\,H,\,A\to\gamma\gamma$ cross sections times branching fractions
as contour plots in the $(m_A,\,\tan\beta)$ plane;
the $H\to \gamma\gamma$ signals are essentially
restricted to the region where $H\to hh$ is forbidden or suppressed (compare
Fig.~\ref{f}). Figure~\ref{m} gives the corresponding
potential discovery regions for $h, H$ or $A$ for luminosity
${\cal L}= 20\,\rm fb^{-1}$ (corresponding approximately to 2~years SSC
running at one intersection). The light areas show where $S\Big/\sqrt B$
exceeds 4 standard deviations.
We note that the region where $h\to\gamma\gamma$ signals are detectable is also
the region where $h$ masquerades as the SM Higgs boson (see Fig.~\ref{c}(a)).

\subsection{$q\bar q \to (h,H)W \to \gamma\gamma\ell\nu$\,;\ \
        $gg,q\bar q\to (h,H,A)t\bar t\to \gamma\gamma\ell\nu\,$jets}

The presence of a lepton tag from $W\to\ell\nu$ or $t\to bW\to b\ell\nu$
decay leads
to a cleaner signal that can be identified in the SDC detector~\cite{sdc}
with $m(\gamma\gamma)$ resolution ${}\alt 3\,$GeV. Simulations show that
 the SM Higgs boson signal can be detected
above the backgrounds (of which $t\bar t\gamma\gamma$ production
is the most important) through the intermediate mass range up to
140\,GeV~\cite{snowmass,sdc,tth}.
Figure~\ref{n} shows the raw signal cross sections from the $t\bar t$ Higgs
process (for the parameters of
Eq.~1) as contour plots, summing over lepton flavors $\ell = e,\mu$.
In the following discussion we focus on detection in SDC.

The $W-$Higgs signals are calculated including an order $\alpha_s$ QCD
enhancement factor\cite{han-willen}.
The $t\bar t$-Higgs signal is evaluated as in
Ref.~\cite{stange-etal} using the scale
$Q = \bigl[m_T(t) + m_T(\bar t) + m_T(h)\bigr]/3$
where $m_T=\sqrt{m^2 + p_T^2}$.
Final leptons and photons are required to satisfy realistic acceptance cuts on
transverse momentum $p_T$, pseudo-rapidity $\eta$ and separation $\Delta R =
\sqrt{(\Delta\phi)^2+(\Delta\eta)^2\,}$
\begin{equation}
\begin{array}{c}
p_T > 20~{\rm GeV}\,, \quad |\eta| < 2.5\,, \\
\Delta R_{\ell\gamma} > 0.4\,, \quad  \Delta R_{\gamma\gamma} > 0.4\,,
\end{array}
\end{equation}
where $\phi$ denotes azimuthal angle.

The cross sections are multiplied by $(0.85)^3$ for the detection
efficiency\cite{sdc} and an additional factor of 0.93 in the $W$-Higgs events
and 0.73 in the $t\bar t$-Higgs events for isolation (excess $\sum E_T < 10$
 GeV in a cone $\Delta R < 0.3$) as determined in SDC simulations.

The Gaussian resolution $\sigma_{\gamma\gamma}$ is taken from the baseline
detector results in Table 3-3 of Ref.~\cite{sdc}
({\it e.g.}\ $\sigma_{\gamma\gamma} = 2$~GeV for $m_H = 160$~GeV).
The $b\bar b\gamma\gamma$ and $W\gamma\gamma$ backgrounds are also taken
from Ref.~\cite{sdc}.  The $t\bar t\gamma\gamma$ background is evaluated
from the formulas for $t\bar t ZZ$ production in Ref.~\cite{stange-etal}, with
appropriate replacement of masses and couplings.  The signals and backgrounds
are integrated over an interval $m_H \pm 2\sigma_{\gamma\gamma}$.
Figure~\ref{o} gives the contours of significance $S\Big/\sqrt B=4$ for the
combined $t\bar t$ Higgs plus $W$ Higgs lepton-tagged signals for
luminosity 20\,fb$^{-1}$.  Light areas show the potential discovery regions
for $h$ or $H$  where the significance exceeds 4~standard deviations;
there are no tagged-$A$ discovery regions.
We see that the use of lepton-tagged two-photon channels substantially
increases the discovery regions (compare Fig.~\ref{m} for untagged cases).

\subsection{$gg\to H\to ZZ\to 4\ell$}

Previous studies\cite{kz,680,gunion} have found that four-lepton
decays are only useful as a signal for $H$; the $h\to ZZ$ branching
fraction is kinematically suppressed and $A\to ZZ$ is absent at tree level.
Figure~\ref{p} displays the $\sigma B$ contours for $pp\to H\to4\ell$
at $\sqrt s=40$\,TeV for the parameters of Eq.~(1), summed over lepton flavors
$\ell = e,\mu$.

The principal background to this process is $q\bar q\to ZZ \to 4\ell$.
We multiply the cross-section for this process by 1.65 to approximate the
contributions of higher order QCD corrections and $gg\to ZZ$
\cite{gunion,bg,bchoz}.  Backgrounds from $t\bar t, Zb\bar b$ and $Zt\bar t$
production are essentially eliminated by our choice of cuts and the
requirement that the four leptons be isolated.  We require that each of the
four leptons satisfy
\begin{equation}
p_{T\ell} > 20~{\rm GeV}\,\qquad  |\eta_\ell| < 2.5\,.
\end{equation}
In addition a separation $\Delta R > 0.4$ is required between all lepton pairs
and an invariant mass restriction
\begin{equation}
|M_Z - M_{\ell\ell}| \le 10~{\rm GeV}
\end{equation}
is imposed on one $\ell^+\ell^-$ pair in the event as expected for one on-shell
$Z$-boson.
We multiply the calculated cross section by $(0.9)^4$ to include SDC
estimates of lepton isolation (excess $\sum E_T < 5$~GeV in a cone
$\Delta R < 0.4$) and multiply by $\left[(0.85)(0.95)\right]^2$ to
include the efficiency for the detection of lepton pairs from $Z$-bosons.

The mass resolution $\sigma_{4\ell}$ of the Higgs boson peak is
estimated using the SDC electromagnetic single particle resolution
\begin{equation}
{\sigma(E) \over E} = \left[\left({0.14\over\sqrt E}\right)^2 + 0.01^2\right]
        ^{1/2}\,.
\end{equation}
We find that $\sigma_{4\ell}$ ranges from $\approx 2$~GeV for
$m_H = 140$~GeV to $\approx 6$~GeV for $m_H = 1000$~GeV.  This resolution
is folded in quadrature with the Higgs boson decay width $\Gamma_H$
to find an effective Gaussian resolution
\begin{equation}
\sigma_{\rm eff} = \left[\sigma_{4\ell}^2 +
        \left({\Gamma_H\over 2}\right)^2\right]^{1/2} \,.
\end{equation}
The signal and backgrounds are integrated over the interval
$m_h \pm 2\sigma_{\rm eff}$ in our analysis.

Figure~\ref{q} gives the resulting contours of $S\Big/\sqrt B=4$
for ${\cal L} = 20\,fb^{-1}$ in the $(m_A,\,\tan\beta)$ plane.
The potential discovery regions with
$S\Big/\sqrt B \ge 4$ and $S \ge 10$ events are unshaded.
As previously remarked~\cite{kz,680,gunion}, this canonical ``gold-plated'' SM
signature for a heavy Higgs boson can only be detected in a limited region of
MSSM parameter space; this is due to competing $H\to hh$ and other decay
modes, not present in the SM ({\it cf.}~Fig.~\ref{f}), and reduced $HZZ$
coupling strength.  The boundary at $m_A \approx 300$~GeV is associated with
the kinematic threshold for $H \rightarrow t\bar t$ decay.  Our realistic
calculations including explicit lepton cuts lead to a smaller potential
discovery region than other recent analyses\cite{kz,gunion}
that assumed an $m_H$-independent detector acceptance.

\subsection{$gg,\,q\bar q\to t\bar t; \; t\to bH^+$}

There will be copious $t\bar t$ production at the LHC and SSC that will
allow non-standard top decays to be scrutinized by triggering on the
$t\to bW^+\to b\ell\nu$ decay of one top quark\cite{tbh-decay}.
Simulations\cite{gem,sdc,roy92} indicate that the decay
$t\to bH^+$ with subsequent $H^+ \to \tau^+\nu$ and
$\tau^+\to\bar\nu_\tau\pi^+$ decays will be detectable as a violation
of lepton universality expected from $W$-decays in
$t\bar t \to W^+W^- \to \ell\nu\tau\nu$ events.
In evaluating this signal we impose the SDC acceptance cuts and
efficiencies~\cite{sdc},
which include a cut $p_T(\pi) > 100$ GeV.  A significance $S\big/ \sqrt{S+B} >
5$ is required for discovery, where $B$ is the number of background
$\ell\nu\tau\nu$ events from SM $t\bar t \to b\bar b WW$ decays and $S$
is the excess of such events due to the charged Higgs decay mode
of the top quark.  Figure~\ref{r} displays the potential discovery regions in
the $(m_A,\tan\beta)$ plane for $m_t = 120$, 150 and 200\,GeV with
${\cal L} = 20$ fb$^{-1}$;  the inaccessible
region for $m_t=150$\,GeV is shaded.  Charged Higgs boson masses up to a few
GeV below $m_t$ are accessible (see Fig.~\ref{a}). We note however that
detecting an excess of $\ell\nu\tau\nu$ events would not by itself measure the
mass $m_{H^+}$, although it would constrain $m_A$ and $\tan\beta$.

\section{SUMMARY AND CONCLUSIONS}

The limits of detectability depend on several different factors:
(a) the $\alpha$- and $\beta$-dependent couplings governing the dynamics
of production and decay, (b) the Higgs boson masses that determine the
regions of kinematical suppression, (c) the acceptance, luminosity
 and background values that determine whether or not a signal can be extracted.
  Of these, (a) and (b) vary strongly across the $(m_A,\,\tan\beta)$
parameter plane and we have evaluated them exactly; (c) can only be
estimated approximately, but moderate changes here will not qualitatively
alter our conclusions about discovery limits in the parameter space.

In the previous sections we have discussed the coverage of the
$(m_A,\,\tan\beta)$ plane achievable by different colliders, using
the best signals in each case. We can now combine these different
discovery limits, to see how completely the parameter space can be explored.

We first discuss the potential for discovery of at least one of the
MSSM Higgs bosons, $h,\,H,\,A,\,H^\pm$.
Figure~\ref{w} shows the combined coverage, assuming $m_t=150$\,GeV
with the SUSY mass scale around 1\,TeV as in Eq.~(1).
Roughly speaking, LEP\,I and LEP\,II can cover all areas except the
region $m_A\agt 80,\ \tan\beta\agt 2$. The SSC/LHC searches for
$h\to\gamma\gamma$ and $H\to\gamma\gamma$ (with or without a lepton tag)
 extend the coverage to include $m_A\agt 180$\,GeV and $m_A\alt 100$\,GeV
respectively (any $\tan\beta$). Charged Higgs searches extend the discovery
region little for this $m_t$ value. The region $100\alt m_A\alt 180$\,GeV  with
$\tan\beta\agt 2$ remains largely inaccessible for the standard luminosities.
However, a higher-energy $e^+e^-$ collider could cover this remaining
area~\cite{janot,barger92}.
The SSC/LHC searches for $H\to\ell\ell\ell\ell$ cover a region already
accessible to other searches ($Zh$ at LEP\,II, $h\to\gamma\gamma$ at SSC/LHC)
but would discover a different particle.
Figure \ref{no} presents an expanded view of the region of the
$(m_A,\tan\beta)$
plane showing the Higgs boson masses that would not be ruled out by
LEP\,II and SSC experiments, for the case $m_t = 150$ GeV.
Figures \ref{x} and \ref{y}  map the discovery regions for the
cases $m_t=120$\,GeV and $m_t=200$\,GeV, respectively. In the case
$m_t=200$\,GeV, the improvement in coverage is largely due to the wider range
of $H^+$ masses that become accessible through $t\to bH^+$ decays.

Finding one MSSM Higgs boson is a minimum requirement. In some parts of
the $(m_A,\,\tan\beta)$ plane, more than one of these Higgs bosons would
be discoverable. This information is summarized in Fig.~\ref{z}, for the
case $m_t=150$\,GeV.

\newpage
\acknowledgements
We thank Micheal Berger and Tim Stelzer for helpful conversations;
we thank Peter Norton for conversations about experimental issues, Paul
Dauncey for information about $b$-tagging efficiencies.
This work is supported in part by the U.S.~Department of Energy under contract
No.~DE-AC02-76ER00881, in part by the Texas National Research Laboratory
Commission under grant No.~RGFY9173, and in part by the University of Wisconsin
Research Committee with funds granted by the Wisconsin Alumni Research
Foundation. Further support was provided by U.S.~Department of Education Award
No.~P200A80214.

%---------------------------------------------------------------------------

%----------------------------------------------------------------------------

\figure{\label{a}
Contour plots of MSSM Higgs boson masses
 in the $(m_A,\,\tan\beta)$ plane, for the SUSY
parameters in Eq.~(1): (a)~$m_h$ and $m_H$ for $m_t=150$\,GeV,
(b)~$m_{H^+}$ for $m_t=150\,$GeV, (c)~$m_h$ and $m_H$ for $m_t=200$\,GeV,
(d)~$m_{H^+}$ for $m_t=200$\,GeV. The $m_h\ (m_H)$ curves are distinguished by
solid (dashed) curves.}

\figure{\label{c}
Contours in the $(m_A,\,\tan\beta)$ plane of the $\alpha$-dependent
factors that enter the $h$ and $H$ couplings to weak bosons and quarks:
(a)~boundaries, to the right of which the $h$ couplings are within 10\% of the
$H_{\rm SM}$ couplings; (b)--(f)~contour plots of the individual coupling
factors $\sin(\beta-\alpha)$, $\cos\alpha/\sin\beta$,
$\sin\alpha/\sin\beta$, $-\sin\alpha/\cos\beta$,
$\cos\alpha/\cos\beta$, respectively.
The $h$ couplings approach the $H_{\rm SM}$ couplings at large $m_A$.}

\figure{\label{d}
The $b$-loop $\left(-\case4/9 I_b\right)$, $t$-loop $\left(-\case16/9
I_t\right)$ and $W$-loop
$\left(7I_W\right)$ factors contributing to the $\gamma\gamma$
partial widths of the $h$ and $H$ Higgs bosons are shown versus
 mass $m_X$.}

\figure{\label{e}
Two-photon partial widths for $h,H,A$ decays versus Higgs boson mass for the
parameters of Eq.~(1) with (a)~$\tan\beta=2$, (b)~$\tan\beta=5$ and
(c)~$\tan\beta=30$.}

\figure{\label{f}
Kinematically allowed regions in the $(m_A,\,\tan\beta)$ plane for
MSSM Higgs boson decays into other Higgs bosons; when allowed, these
decays usually dominate over other decay modes. Shading shows the area where
the new modes $H\to hh,AA,AZ$ are all forbidden. A dashed curve shows the locus
of zeros of the $H\to hh$ coupling.}

\figure{\label{g}
Total widths of MSSM Higgs boson decays versus mass compared to the
SM result, for (a) $\tan\beta=2$, (b) $\tan\beta=30$, with the
parameters of Eq.~(1).}

\figure{\label{new8}
Contour plots of total widths of MSSM Higgs bosons in the $(m_A,\,\tan\beta)$
plane for (a)~$h$ decay, (b)~$H$ decay, (c)~$A$ decay, (d)~$H^+$ decay,
with the parameters of Eq.~(1).}

\figure{\label{h}
MSSM Higgs boson branching fractions for the dominant modes and the
most detectable modes, taking the parameters of Eq.~(1):
(a) $h,H$ decays with $\tan\beta=2$, (b) $h,H$ decays with $\tan\beta=30$,
(c) $A$ decays with $\tan\beta=2$, (d) $A$ decays with $\tan\beta=30$,
(e) $H^+$ decays with $\tan\beta=2$, (f) $H^+$ decays with $\tan\beta=30$.}

\figure{\label{i}
Excluded regions of the $(m_A,\,\tan\beta)$ plane, from the ALEPH
experiment~\cite{aleph} at LEP\,I, including electroweak radiative
corrections with the SUSY parameters of Eq.~(1) and
$m_t=100$, 120, 150 or 200~GeV:
(a)~$Z\to Z^*h$ searches,
(b)~$Z\to Ah$ searches.
Areas below or to the left of the curves are excluded; the non-excluded
regions for $m_t = 150$ GeV are shaded.}

\figure{\label{j}
Limits of detectability at LEP\,II for the $e^+e^-\to Zh$ signals
obtained by rescaling the simulations of Ref.~\cite{saulan},
for luminosity 0.5\,fb$^{-1}$ and the SUSY parameters of Eq.~(1) with
$m_t=100$,
120, 150 or 200~GeV. The inaccessible area for $m_t=150\,$GeV is shaded.}

\figure{\label{k}
Regions of the $(m_A,\,\tan\beta)$ plane where an MSSM Higgs boson signal
would be detectable in the $\tau\tau jj$ channel at LEP\,II, for
luminosity 0.5\,fb$^{-1}$ and the SUSY parameters of Eq.~(1) with $m_t= 120$,
150 or 200~GeV. Boundary curves are shown for different top masses; the area
inaccessible with $m_t=150$\,GeV is shaded.}

\figure{\label{l}
Untagged two-photon signals at the SSC.
Predicted cross sections times branching fraction in fb at $\sqrt s=40$\,TeV
for $pp\to h,H,A\to \gamma\gamma$ signals are shown as contour plots
in the $(m_A,\,\tan\beta)$ plane, for the parameter choices of Eq.~(1):
(a) $h$ signals, (b) $H$ signals, (c) $A$ signals.}

\figure{\label{m}
Contours of significance $S\Big/\sqrt B=4$ for the detection of the untagged
two-photon signals in
Fig.~\ref{l}, based on a luminosity of 20\,fb$^{-1}$ and the GEM
estimates of backgrounds~\cite{gem}. The light areas show potential discovery
regions for $h, H$ or $A$, where the significance exceeds four standard
deviations.}

\figure{\label{n}
Lepton-tagged two-photon signals at the SSC.
The cross sections times branching fraction in fb at
$\sqrt s=40$\,TeV for the  process
$pp\to gg,q\bar q\to (h,H,A)t\bar t\to \gamma\gamma\ell\nu$\,jets
are shown as contour
plots in the $(m_A,\,\tan\beta)$ plane, for the parameter choices of
Eq.~(1): (a) $h$ signals, (b) $H$ signals, (c) $A$ signals.}

\figure{\label{o}
Contours of significance $S\Big/\sqrt B=4$ for detection of the combined $t\bar
t$ Higgs and $W$ Higgs lepton-tagged two-photon signals
in Fig.~\ref{n}, based on luminosity 20\,fb$^{-1}$ with SDC estimates of $b\bar
b\gamma\gamma$ and $W\gamma\gamma$ backgrounds and our calculations
of the $t\bar t \gamma\gamma$ background, integrated over 3~GeV bins
of $\gamma\gamma$ invariant mass. The light areas show potential discovery
regions for $h$ and $H$.}

\figure{\label{p}
Four-lepton signals at the SSC.
Cross sections times branching fraction $\sigma B(pp\to H\to 4\ell)$
 at the SSC are shown in fb, as contours in the $(m_A,\,\tan\beta)$ plane, for
the parameters of Eq.~(1).}

\figure{\label{q}
Contours of significance $S\Big/\sqrt B=4$ for detection of the four-lepton
signals in Fig.~\ref{p}, based on luminosity 20\,fb$^{-1}$ and calculated
$ZZ\to4\ell$ background. The light areas show potential $H$ discovery regions.}

\figure{\label{r}
Potential discovery regions for the charged Higgs boson via the decays
$t \to bH^+$, $H^+ \to \tau^+\nu$, $\tau^+ \to \pi^+\nu$, through lepton
universality violation in $t\bar t \to W^+W^- \to \ell\nu\tau\nu$ events;
the inaccessible region for $m_t = 150$ GeV is shaded.}

\figure{\label{w}
Combined potential discovery regions for at least one MSSM Higgs boson, with
$m_t=150$\,GeV and the SUSY parameters of Eq.~(1), are shown as an unshaded
area in the $(m_A,\,\tan\beta)$ plane. Curves show the limits of various search
possibilities, previously described in Figs.~\ref{i}--\ref{r}: (a)~LEP\,I and
LEP\,II limits, (b)~SSC limits.}

\figure{\label{no}
Expanded view of the region of no coverage from Fig.~\ref{w}, for
$m_t=150$\,GeV, shown as a shaded area in the $(m_A,\,\tan\beta)$ plane.
Contours of $m_h$ (solid), $m_H$ (dashed) and $m_{H^+}$ (dotted) are
superposed.}

\figure{\label{x}
Similar to Fig.~\ref{w} with $m_t=100$\,GeV.}

\figure{\label{y}
Similar to Fig.~\ref{x} with $m_t=200$\,GeV.}

\figure{\label{z}
Potential discovery regions for more than one MSSM Higgs boson, with
$m_t=150$\,GeV and the SUSY parameters of Eq.~(1). The regions where no bosons,
one boson and two or more bosons can be detected are distinguished by different
levels of shading.}

\end{document}